\documentclass[aps,prb,twocolumn,superscriptaddress,epsfig,floats,showpacs]{revtex4}
\usepackage{graphicx}
\usepackage{amssymb,bm}
\usepackage{epstopdf}
\usepackage{color}
\usepackage{amssymb,amsmath}
\begin{document}

\title{Kondo effect in an antiferromagnetic metal}
\author{Vivek Aji}
\affiliation{Department of Physics and Astronomy, University of
California, Riverside, CA 92521}\author{Chandra M. Varma}
\affiliation{Department of Physics and Astronomy, University of
California, Riverside, CA 92521}
\author{Ilya Vekhter}
\affiliation{Department of Physics and Astronomy, Louisiana State
University, Baton Rouge LA 70803}
\begin{abstract}
We study the fate of a spin-1/2 impurity in the itinerant
antiferromagnetic metallic phase via a renormalization group
analysis and a variational calculation. The local moment -
conduction electron interaction hamiltonian in an
antiferromagnetic metal is spin non-conserving. We show that for a
general location of the impurity, the Kondo singularities still
occur, but the ground state has a partially unscreened moment. We
calculate the magnitude of this residual moment and the variation
of the spin polarization with energy for a substitutional impurity
as a function of the staggered magnetization. The usual Kondo
effect only occurs if the impurity is placed at points where the
magnetization is zero.
\end{abstract}

 \maketitle

\section{Introduction.}

The dual nature of electron behavior in antiferromagnetic (AFM)
heavy fermion (HF) materials is not well understood. On one hand,
in many intermetallic compounds containing elements with partially
filled f-orbitals, the heavy mass originates from the collective
screening of the local moments by the conduction band via the
Kondo effect (for a review see Refs.\onlinecite{ACH,OF,CMV1}).
Emergence of magnetic order, on the other hand, requires
interaction between the unscreened moments.

In many materials exhibiting coexistence of antiferromagnetism and
heavy mass the RKKY interaction between the local moments is
comparable to the Kondo temperature \cite{DON,CMV},  and the AFM
order involves only {\it part} of the local moments at low
temperatures \cite{CB}. This suggests a ground state where a
fraction of the full unscreened moment appears as ordered
staggered moment, while the rest continues to be compensated by
the conduction electrons via the Kondo effect \cite{CB}. In
materials with several $f$-electrons per unit cell, such as
several U-containing compounds, the coexistence of AFM and HF
behavior may be understood as originating from Kondo screening of
one $f$-species and ordering of another. In systems with a single
$f$-electron, for example many Ce-containing intermetallics, the
coexistence of magnetic ordering and screening originates with the
same local moment.

Motivated by this picture we consider in this paper the Kondo
effect for a single impurity coupled to a band of conduction
electrons that order antiferromagnetically. We believe that this
is the first step towards developing a low-energy theory for the
coexistence of the AFM and HF behavior. Just as the essential
aspects of singlet formation and mass enhancement in paramagnetic
heavy fermions follow from the analysis of the screening of a
local moment in an electron bath, the salient features of
competition between antiferromagnetism and Kondo screening can be
understood from the analysis of our model. We write down a simple
model hamiltonian for this problem and show that in a mean-field
theory for the lattice, our model has precisely the local
symmetries required to consider the competition between the two
phenomena. We find that only partial screening of the impurity
spin takes place due to spin non-conserving interaction vertices,
with the unscreened moment fraction that depends on the amplitude
of the AFM order.

A major advantage of our model is that it allows a variety of
analytic approaches that make the underlying physical picture more
transparent. Previous analyses of the competition between
antiferromagnetism and screening for the Kondo lattice or
equivalent models have been largely numerical. Several authors
considered a half-filled Kondo lattice with anisotropic Kondo
coupling~\cite{GMZhang:2000,SCapponi:2001}. In that case the AFM
phase is insulating, rather than metallic, in contrast to the
experimental situation. Moreover, the coexistence regime does not
exist for the isotropic Kondo coupling that we consider, although
it may appear under applied magnetic fields~\cite{KBeach:2004}. A
metallic phase with coexisting Kondo screening and AFM order was
recently found in variational Monte Carlo calculations for the
Kondo Lattice Model~\cite{HWatanabe:2007}, however, Kondo
screening was argued to remain the same in the paramagnetic and
antiferromagnetic phases, in contrast to our findings below.

Our study also has fundamental importance beyond the connection to
the heavy fermion physics. Our main conclusions, (i) that the
Kondo screening of an impurity placed in an itinerant
antiferromagnet (a) depends on the location of the impurity within
the unit cell; (b) for substitutional impurity on a simple lattice
is incomplete with the magnitude of the residual moment related to
the amplitude of the AFM order; (ii) that the spin-dependent local
density of states reflects this incomplete screening can all be
tested experimentally. It is important to emphasize that, for a
system without nesting, after the onset of the AFM order and
doubling of the unit cell there remains a large Fermi surface with
non-vanishing density of states, and one may naively expect a
complete Kondo screening. The existence of residual unscreened
moment is due to the spin non-conserving vertices in the
magnetically ordered state that alter the interaction between the
conduction electron bath and the local moment.

The remainder of the paper is organized as follows. In the next section we introduce our model and discuss its salient features. The structure of the interaction vertices and the renormalization
group analysis are presented in Sec.\ref{sec:RG}. We then present the variational ground state ansatz in Sec.\ref{sec:var}, and follow it by the discussion of the density of states on the
impurity site.

\section{Model.}
\label{sec:model}

Our model consists of an electronic band with an itinerant AFM
modulation coupled to an isolated impurity, with the hamiltonian
$H=H_{c}+H_{I}+H_K$. The conduction band without magnetic order is
described by
\begin{equation}
  H_{c} = \sum_{{\bf k},\alpha}\varepsilon
\left({\bf k}\right) c_{{\bf k},\alpha}^{\dag} c_{{\bf
k},\alpha}\,,
    \label{eq:PM}
\end{equation}
where $\varepsilon\left({\bf k}\right)$ is the energy band
dispersion, and $c_{{\bf k},\alpha}^{\dag}$ is the creation
operator for an electron eigenstate in a paramagnet. The AFM order
is imposed via
\begin{equation}
  H_{I}= \sum_{{\bf k}}\left[\gamma m
\left(c_{{\bf k},\uparrow}^{\dag} c_{{\bf k}  +
\textbf{Q},\uparrow} - c_{{\bf k},\downarrow}^{\dag} c_{{\bf
k}+\textbf{Q},\downarrow}\right)  + h.c. \right]\,,
    \label{eq:AFM}
\end{equation}
where $m$ is the mean field staggered magnetization, chosen along
the spin $z$-axis, $\gamma m$ is the AFM gap, and $\textbf{Q}$ is
the ordering wave-vector.

We consider an isolated impurity with spin $\bm S$, at position
${\bf R}_i$ coupled antiferromagnetically to the fermion spins.
The Kondo interaction is
\begin{eqnarray}
H_K=- J {\bf S}_i\cdot {\bm\sigma}_{\alpha\beta}
\psi^\dagger_\alpha ({\bf R}_i)\psi_\beta ({\bf R}_i)\,,
    \label{Eq:Ham}
\end{eqnarray}
\noindent
    where $J$ is the (Kondo) exchange coupling
between the impurity spin and the conduction electrons,
$\bm\sigma$ is the vector of Pauli matrices, and
$\psi^\dagger_\alpha(\bm R)=\sum_{\bf k} c_{{\bf
k},\alpha}^{\dag}\exp(i{\bf k}\cdot\bm R)$, is the creation
operator in real space.

The justification for this model is as follows. Suppose we
consider the full Kondo lattice Hamiltonian including the RKKY
interaction and allow for the possibility of antiferromagnetic
order so that the local spin at each site may be written as
\begin{equation}
  {\bf S}_i = \langle{\bf S}_{\bf Q}\rangle\exp({i{\bf Q\cdot
R}_i}) + \left[{\bf S}_i -\langle{\bf S}\rangle_{{\bf
Q}}\exp({i{\bf Q\cdot R}_i})\right].
\end{equation}
Now consider a mean field
theory (for example the dynamical mean-field theory). At the first
iteration of the self-consistent solution, one must solve the
local "impurity" problem, without making this decomposition at the
impurity site but making it everywhere else to generate a
self-consistent bath. Then Eq.(\ref{eq:AFM}) follows as the
periodic spin-dependent potential on the conduction electrons due
to the staggered magnetization, $\gamma m\propto J\langle \bm
S_{\bm Q}\rangle$. Therefore the local problem is precisely the
problem with the symmetries of Eqs.(\ref{eq:PM})-(\ref{Eq:Ham}).
For a fully self-consistent solution we would have to relate
$\gamma m$ to the sub-lattice moment generated at the impurity
site(s) due to it; this second part is not attempted here. The
self-consistency affects the quantitative details, but not
qualitative physics of our conclusions. Our finding that there is
only partial Kondo screening in the local model confirms the
consistency of the approach.

In the absence of the Kondo coupling, the Hamiltonian,
Eqs.(\ref{eq:PM})-(\ref{eq:AFM}) can be readily diagonalized using
a Bogoliubov transformation. This introduces four species of
fermions, described by the band indices $n=\pm$, in addition to
the two spin projections on the $z$-axis $\alpha=\pm 1$.
\begin{eqnarray}
\begin{pmatrix} a_{+,{\bf k},\alpha}^{\dag}\\ a_{-,{\bf
k},\alpha}^{\dag} \end{pmatrix}&=& \begin{pmatrix} u_{{\bf
k}}c_{{\bf k},\alpha}^{\dag}-\alpha v_{{\bf k}}c_{{\bf
k}+\textbf{Q},\alpha}^{\dag}\\ v_{{\bf k}}c_{{\bf
k},\alpha}^{\dag}+ u_{{\bf k}}c_{{\bf
k}+\textbf{Q},\alpha}^{\dag}\end{pmatrix}\,,
\end{eqnarray}
with the Bogoliubov factors
\begin{equation}
  \left\{u_{{\bf k}}^{2}\,, v_{{\bf k}}^{2} \right\}=
\frac{1}{2}\left[ 1 \pm {\delta_{{\bf k}}/ {\sqrt{\delta_{{\bf
k}}^{2} + \gamma^{2}m^{2}}}}\right]\,.
\end{equation}
Under
this transformation the band hamiltonian becomes
\begin{equation}
  H_b=H_c+H_I=\sum_{{\bf k}, n , \alpha}
E_{n}({\bf k})a^{\dag}_{n{\bf k}\alpha} a_{n{\bf k}\alpha}\,,
    \label{eq:aBand}
\end{equation}
with the energy dispersion $ E_{\pm}({\bf k}) = \xi_{{\bf k}} \pm
\sqrt{\delta_{{\bf k}}^{2} + \gamma^{2}m^{2}}$, and the shorthand
notation $\xi_{{\bf k}} = {1\over 2}\left( \varepsilon ({\bf k}) +
\varepsilon ({\bf k} + \textbf{Q}) \right)$ and $\delta_{{\bf k}}
= {1\over 2}\left(\varepsilon ({\bf k}) - \varepsilon ({\bf k} +
\textbf{Q}) \right)$. The momentum summation in
Eq.~(\ref{eq:aBand}) is over the first magnetic Brillouin Zone
(MBZ). In the absence of nesting, $\xi_{\bf k}\neq 0$, the
electrons remain ungapped, and the MBZ contains a Fermi surface
(FS). Superficially, since the density of states at the Fermi
level is finite, this may suggest that Kondo screening should
occur as in a normal metal. We show below, however, that this is
not the case since the Kondo interaction vertices have nontrivial
spin structure.

The effective hamiltonian is now written in terms of the operators
$a_{n,{\bf k},\alpha}$ as $H = H_{b} + H_{K}$ where $H_b$ is given
by Eq.~(\ref{eq:aBand}), and the Kondo interaction is
\begin{subequations}
\begin{eqnarray}
&&H_{K}=-{J\over {2N}}\sum_{{\bf k}{\bf k}^{'}\atop
nn'}\textbf{S}\cdot {\bm\sigma}_{\alpha\beta}f_{n{\bf
k}\alpha}f^\star_{n'{\bf k}^{'}\beta} a^{\dag}_{n{\bf k}\alpha}
a_{n'{\bf k}^{'}\beta}\,,\label{eq:aKondo}
\\
 && f_{+\bm k  \alpha}= {u_{\bf k}} -\alpha{v_{\bf k}} e^{i\bm Q{\bf
  R}_i}\,,
  \\
 && f_{-\bm k\alpha}=\alpha{v_{\bf k}} +{u_{\bf k}} e^{i\bm Q{\bf
  R}_i}\,.
\end{eqnarray}\end{subequations}
Again, the momentum summation in Eq.~(\ref{eq:aKondo}) is over the
first MBZ. Eq.(\ref{eq:aKondo}) shows that the effective Kondo
interaction, written in terms of the eigenstates of the
antiferromagnet is, in general, anisotropic, {\bf k}-dependent, as
well as dependent on the location of the impurity in the lattice
because of the structure factors, $f_{n{\bf k}\alpha}$. The
impurity spin is coupled to fermions in both bands, and this
interaction is weighted by the structure factors. Despite this
apparent complication, we show below that the invariant vertices
have a factorizable form so that the problem is still tractable.

\section{Renormalization group analysis}
\label{sec:RG}

\subsection{Spin structure of invariant vertices} 
\begin{figure}[t]
  \begin{center}
  \includegraphics[width=0.4\columnwidth]{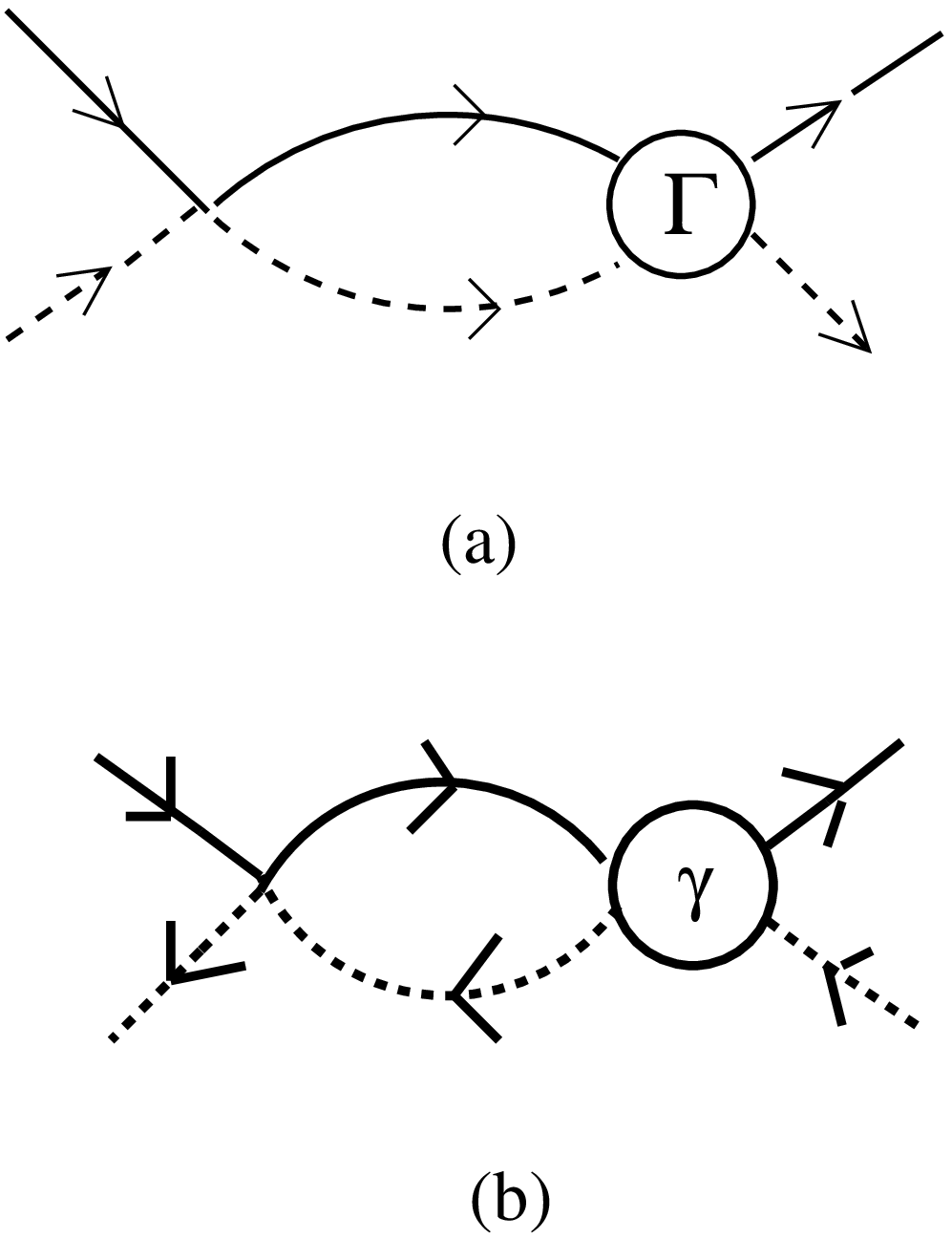}
  \hspace*{0.25cm}
  \includegraphics[width=0.4\columnwidth]{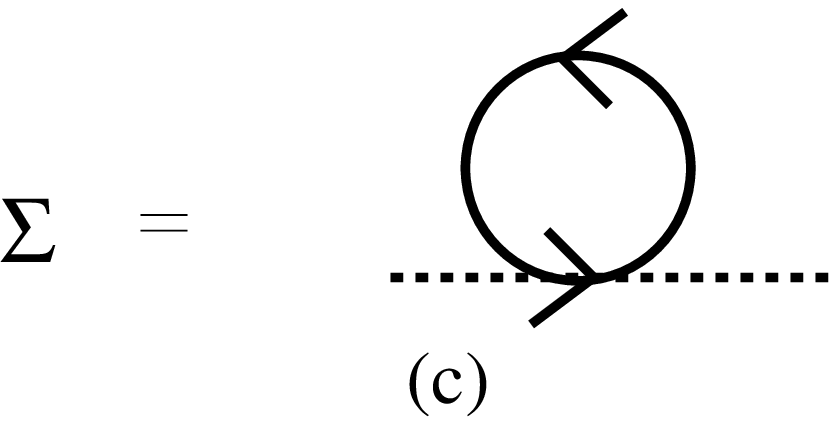}
  \caption{a) Particle-Particle channel, (b) Particle-hole
  channel contributions to the vertex and (c) Contribution to the Pseudo-Fermion Self Energy }
  \label{fig:selfen}
  \end{center}
\end{figure}

To identify the
invariant vertices we employ the Abrikosov pseudo-fermion
representation for the local moment, ${\bf S}=\sum_{\mu\nu}{\bm
S}_{\mu\nu}\psi^\dagger_\mu \psi_\nu$, where ${\bm S}_{\mu\nu}$ is
the spin matrix~\cite{ABR}. The integral equation for the
interaction vertex is represented in Fig.~\ref{fig:selfen} and
contains contributions from the particle-particle
(Fig.~\ref{fig:selfen}$(a)$) and the particle-hole
(Fig.~\ref{fig:selfen}$(b)$) channels. For
Fig.~\ref{fig:selfen}(a) we have ($k_B=\hbar=1$)

\begin{widetext}
\begin{eqnarray}
\Gamma^{\alpha\beta n}_{\alpha^{'}\beta{'} n^{'}}
        ({\bf k}, {\bf k'})&=& {J\over {2N}}
\left(\textbf{S}_{\beta\beta^{'}}\cdot\bm{\sigma}_{\alpha\alpha^{'}}\right)
 f_{n{\bf k}\alpha}f^\star_{n^{'}{\bf k}^{'}\alpha^{'}}
 \label{eq:vertex}
 \\
    \nonumber &+&
    {J \over {2N}}\ T \sum_{\alpha^{''}n^{''}  \beta^{''}{\bf q}\ \omega_s}
    \left(\textbf{S}_{\beta\beta^{''}} \cdot
    \bm{\sigma}_{\alpha\alpha^{''}}\right)
    f_{n{\bf k}\alpha}f^\star_{n^{''}{\bf q} \alpha^{''}}
    G_{n^{''}\alpha^{''}}\left(\imath \omega_{s},{\bf{q}} \right)
    F_{\beta^{''}}\left(-\imath \omega_{s}+\imath \omega\right)
\Gamma^{\alpha^{''}\beta^{''} n^{''}}_{\alpha^{'}\beta{'} n^{'}}
 ({\bf q},{\bf k'})
    \,,
\end{eqnarray}
\noindent where $G$  and $F$ are the band fermion and the
pseudofermion propagators respectively. The contribution of
particle-hole channel is obtained by changing the sign of the
frequency in one of the fermion propagators.

We find the solution for the vertex in the factorized form
$\Gamma^{\alpha\beta n}_{\alpha^{'}\beta{'} n^{'}}
        ({\bf k}, {\bf k'})= \Gamma_{\alpha\alpha';\beta\beta'} f_{n{\bf k}\alpha}f^\star_{n^{'}{\bf
k}^{'}\alpha^{'}}$.  The spin matrix
$\Gamma_{\alpha\alpha';\beta\beta'}$ is separated into distinct
symmetry channels which couple the impurity moment to the
fermions. From Eq.~(\ref{eq:vertex}) we find
\begin{eqnarray}
\Gamma^{T}_{\alpha\alpha';\beta\beta'} &=&
\Gamma_{0}^{T}\left(\delta_{\alpha\alpha^{'}}S^{0}_{\beta\beta^{'}}\right)
+\Gamma_{1}^{T}\left(\bm{\sigma}_{\alpha\alpha^{'}}\cdot\textbf{S}_{\beta\beta^{'}}\right)
+
\left(\Gamma_{2}^{T}+\Gamma_{3}^{T}\right)\sigma^{0}_{\alpha\alpha^{'}}S^{z}_{\beta\beta^{'}}
-\left(\Gamma_{2}^{T}-\Gamma_{3}^{T}\right)\sigma^{z}_{\alpha\alpha^{'}}S^{0}_{\beta\beta^{'}}
    \\ \nonumber && \qquad +\
\imath\Gamma_{4}^{T}\left(\sigma^{x}_{\alpha\alpha^{'}}S^{y}_{\beta\beta^{'}}
-\sigma^{y}_{\alpha\alpha^{'}}S^{x}_{\beta\beta^{'}}\right)
+\Gamma_{5}^{T}\left(\sigma^{z}_{\alpha\alpha^{'}}S^{z}_{\beta\beta^{'}}\right),
\end{eqnarray}
\end{widetext}
where the label $T$ refers to the sum of the particle-hole and
particle-particle channels. In the equation above, the vertices
$\Gamma_1$ and $\Gamma_5$ conserve the total spin, $\bm
S+\bm\sigma$, of the impurity-conduction electron system, while
the vertices $\Gamma_2$, $\Gamma_3$, and $\Gamma_4$ do not. This
spin non-conservation is the main reason for the nontrivial
results found below.

\subsection{Impurity location.} The structure of the vertex depends on the
location of the impurity. Consider, for example, a system with a
square lattice, and  the AFM ordering wave vector ${\bf
Q}=(\pi,\pi)$. If the impurity is located at the point where the
magnetization vanishes, midway between two nearest neighbors
(${\bf R}_i=(1/2, 0)$ or $(0,1/2)$),
$\exp\left(\textbf{Q}\cdot\textbf{r}_{I}\right)=\imath$.  Then
$f_{{\bf k}n\alpha}^{\ast}f_{{\bf k}n\alpha}=1$ and no change in
Kondo screening occurs on going from the paramagnetic to the AFM
state. This special behavior takes place only if the impurity is
located at the point where the magnetization vanishes. For a
general position of impurity $f_{n{\bf
k}\alpha}^{\ast}f^\star_{n{\bf k}\alpha}\neq 1 $, so that finite
spin non-conserving vertices exist. In the following, we consider
a substitutional impurity, ${\bf R}=0$.

\subsection{One-loop Renormalization Group Analysis}  
\begin{figure}[t]
  \includegraphics [width=0.8\columnwidth,angle=0]{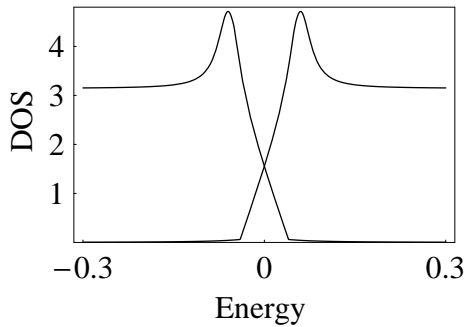}
  \caption{ Density of states for the two bands. For all numerical calculations, except when noted otherwise,
   the bandwidth is taken
  to be 10 and $\gamma m=0.05$.}
  \label{fig:dens}
\end{figure}

We analyze
the equations for the vertices at the one loop level, replacing
the full propagators in Eq.(\ref{eq:vertex}) by their bare
counterparts, $G^0$ and $F^0$. First note the spin-dependent
elastic contribution to the pseudo-fermion self energy, shown in
Fig.~\ref{fig:selfen}(c) $ \Sigma_{f} (\sigma) \sim \sigma \rho
J_{z} m \sinh^{-1}({-D/ \gamma {m}})$, where $\rho$ is the density
of states (DOS), $D$ is the bandwidth, and we assumed a constant
DOS, $N_0$, in the band. Since we assume the Fermi surface is not
nested, opening of the gap at parts of the FS close to the
magnetic Brillouin Zone boundary leads to a build up of density of
states, $D_{\pm} (\epsilon)$ for the '$+$' and '$-$' bands
respectively, at the gap edge $E_{\pm}({\bf k}) \sim \gamma\left|
m \right| $. The detailed shape of the DOS depends on the
underlying band structure of the metal. Here, for simplicity and
without loss of generality, we model $D_{\pm}
(\epsilon)=D_1(\epsilon)+D_{2\pm}(\epsilon)$, where
$D_1(\epsilon)=N_0 \mbox{min}(|\epsilon|/\gamma m, 1)$ is the
linearly suppressed DOS below the AF gap edge, and the Lorentzian
contribution, $D_{2\pm}(\epsilon)=(a\gamma
m/\pi)\left[(\epsilon\mp\gamma m)^2+(\gamma m)^2\right]^{-1}$. For
weak AFM, $\gamma m\ll D$, the choice $a=\gamma m N_0 /2$
conserves the number of states. The resulting DOS for the
particle-hole symmetric case, $D_{-}(-\epsilon)=D_{+}(\epsilon)$,
is shown in Fig.~\ref{fig:dens}: most of the lower(upper) band is
occupied(empty). Despite this complex behavior of the density of
states, evaluating $\Sigma_{f} (\sigma)$ with constant DOS is
qualitatively accurate since there is no (logarithmic) frequency
dependence to the self energy at this order. We use the full model
density of states in computing the renormalization group flow and
carrying out the variational ansatz below.

To one loop we find that $\Gamma^{T}_{2}=-\Gamma^{T}_{3}$, while
to $O(J^{2})$ $\Gamma^{T}_{4}=0$. Both in the particle-hole and
particle-particle channels the terms $\sigma^{0}S^{z}$ and
$\sigma^{z}S^{0}$ appear with opposite signs simply due to spin
commutation rules. However, if the impurity is at a site with
non-vanishing local magnetization, an additional relative sign
change for the $\sigma^{z}S^{0}$ term between the two diagrams in
Fig.~\ref{fig:selfen} is due to the splitting of the band fermion
spin states at the location of the impurity; hence the
contributions of the two diagrams add in this channel, while
canceling in the $\sigma^{0}S^{z}$ channel. To this order in
perturbation theory, the spin dependent self energy of the
pseudofermions is zero, and hence
$\Gamma^{T}_{2}=-\Gamma^{T}_{3}$; this is not true at higher
orders.  The particle-particle and particle-hole contributions to
$\Gamma_{4}$ cancel to all orders due to particle-hole symmetry in
this channel (recall that the magnetization $m$ is chosen along
$z$). In the absence of particle-hole symmetry, however, this spin
non-conserving term is a relevant perturbation about the ordinary
Kondo effect. With particle-hole symmetry, $\Gamma^{T}_{2}$ and
$\Gamma^{T}_{3}$ are the only such relevant perturbations, while
the asymmetry introduced through $\Gamma^{T}_{5}$ is only marginal
at the fixed point.

\subsection{Renormalization Group (RG) Flow.} We now consider the
renormalization of the coupling constants by integrating out the
fermion states near the band edge. As explained above, the spin
non-conserving $\sigma^z S^0$ interaction channel appears in the
vertex. To investigate its flow,  we introduce an effective
exchange field acting only on the fermions: even though such an
interaction is absent from the starting Hamiltonian, it is
generated at one loop level and determines the RG flow by
polarizing the conduction band at the impurity location. To this
end we add to the Hamiltonian the term
    \begin{equation}
    H_{V} = -{v\over {2N}} \sum_{{\bf k},{\bf k}^{'},n,n'}S^0
{\sigma}^{z}_{\alpha\beta}f_{n{\bf k}\alpha}f^\star_{n'{\bf
k}^{'}\beta} a^{\dag}_{n{\bf k}\alpha} a_{n'{\bf k}^{'}\beta}\,.
    \end{equation}
In solving the flow equation we impose the initial condition that
$v=0$ and follow its evolution. Defining $h\left(\omega\right) =
\gamma m/\sqrt{\omega^{2} + \gamma^{2}m^{2}}$, we find the RG
equations for the coupling constants to order $O(J^{2}, v^{2}, J v
)$
\begin{eqnarray}
 && \hspace*{-1cm}{\delta J_{z}\over {2N}}  =
-\left(f(\omega){J_{\bot}^{2}\over {(2N)^{2}}} -g(\omega){v
J_{z}\over {(2N)^{2}}}h\left(\omega\right)\right)
{\delta\omega\over \omega}\,,
    \\
&& \hspace*{-1cm} {\delta J_{\bot}\over {2N}} = - f(\omega){J_{z}
J_{\bot}\over{(2N)^{2}}}{\delta\omega \over \omega}+ g(\omega){(v)
J_{\bot}\over{2(2N)^{2}}}{ h\left(\omega\right) \delta\omega\over
\omega }\,,
    \\
    && \hspace*{-1cm}
{\delta v \over (2N)}= g(\omega)\left({{v}^{2}\over {(2N)^{2}}}+
{y\over {(2N)^{2}}}\right){ h\left(\omega\right)\delta\omega \over
{\omega}}\,,
\end{eqnarray}
\noindent
where,
    \begin{eqnarray}
    \nonumber f(\omega)  &=& (D_{+}(\omega)+D_{-}(-\omega)+D_{+}(-\omega)
+D_{-}(\omega)) \,,\\
\nonumber
    g(\omega) &=&
(D_{+}(\omega)+D_{-}(-\omega)-D_{+}(-\omega)-D_{-}(\omega))\,,
    \end{eqnarray}
 and $
y = {J_{z}^{2}\over 2} - J_{\bot}^{2}$. Note that for a
paramagnetic metal with particle-hole symmetry $g(\omega)=0$, and
the equations reduce to those for the usual Kondo effect
\cite{ACH}.

\begin{figure}[t]
  \includegraphics[width=\columnwidth]{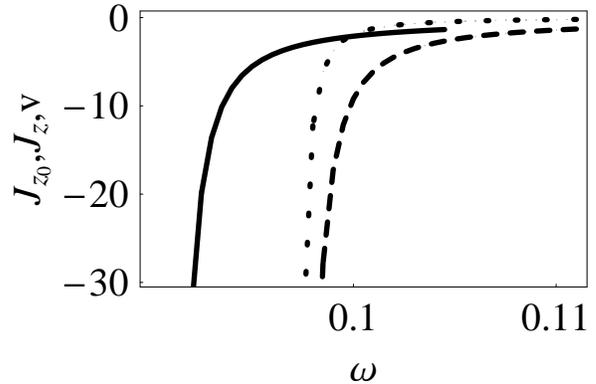}
  \caption{ Flow of $v (\cdot \cdot)$ and $J_{z}(---)$ where the initial values are $J_{z}=-0.4$ and
  $J_{\bot}=-0.4$. For comparison we also show the flow of $J_{z_{0}}(-)$in the absence of AFM order.}
  \label{fig:result1}
\end{figure}

The flow of the exchange field $v$ and the coupling constant
$J_{z}$ is shown in Fig.~\ref{fig:result1} for the density of
states from Fig.~\ref{fig:dens}. For comparison we also show the
flow of $J_{z_{0}}$ in the absence of antiferromagnetic order.
Note that both the Kondo coupling $J$, and the vertex $v$ grow,
and, at the one-loop level (with no self energy corrections), flow
to strong coupling.  This indicates that the Kondo screening is
modified in the AFM state. To understand the nature of the fixed
point we perform a calculation using a variational wavefunction
motivated by the one loop results.

\section{Variational Ansatz.}
\label{sec:var}

To estimate the moment, binding energy and the impurity density of
states in the ground state, we diagonalize the Hamiltonian within
a restricted subspace spanned by the states ($\Downarrow$ and
$\Uparrow$ are the states of the impurity spin, and $n=\pm$ as
before)
\begin{eqnarray}
\nonumber \psi_{1n{\bf k}} &=&\frac{1}{\sqrt 2}\left[\left(u_{{\bf
k}}-nv_{{\bf k}}\right)a_{n {\bf
k}\uparrow}^{\dagger}\otimes\Downarrow - \left(u_{{\bf k}}+n v_{
{\bf k}}\right)a_{{\bf
k}\downarrow}^{\dagger}\otimes\Uparrow\right]\,,
    \\
    \nonumber
\psi_{2n{\bf k}} &=&\frac{1}{\sqrt 2} \left[\left(u_{{\bf k}}+n
v_{{\bf k}}\right)a_{n {\bf k}\uparrow}^{\dagger}\otimes\Downarrow
+ \left(u_{{\bf k}}-n v_{{\bf k}}\right)a_{n{\bf
k}\downarrow}^{\dagger}\otimes\Uparrow\right]\,.
    \end{eqnarray}

The states chosen here generalize those used to describe the
ground state in the standard Kondo problem \cite{VY}, which yield
the same result for the ground state as the large N
approximation~\cite{TVR} or slave Boson method~\cite{PC}.  In
particular, the state $\psi_{1}$ for $u_{\bf k}=1, v_{\bf k}=0$
reduces to the singlet formed between the conduction electron and
the impurity spin, while $\psi_{2}$ in the same limit becomes the
triplet component. In contrast to the paramagnetic metal, here
these two states are coupled by the exchange interaction, and
hence both need to be considered when constructing the variational
wave function.

\begin{figure}[t]
  \includegraphics[width=0.75\columnwidth]{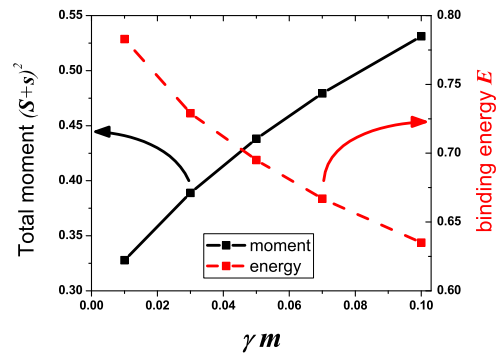}
  \caption{Binding energy normalized to that in the absence of magnetic order, $E_{0}(\gamma m=0)$
  and net moment of the collective state for fixed value of the exchange coupling, $J=-0.04$.}
  \label{fig:bin}
\end{figure}

We approximate $u_{{\bf k}}$ and $v_{{\bf k}}$ for all momentum
states  with the same energy, $\epsilon$, by $u_{\epsilon}$ and
$v_{\epsilon}$, and work in basis states labeled by energy.
Diagonalization within this subspace gives the eigenstates in the
form: $ \psi =
\sum_{\tau\epsilon}\left(a\left(\epsilon,\tau\right)\psi_{1\tau\epsilon}+b\left(\epsilon,\tau\right)
\psi_{2\tau\epsilon}\right) $. We use the density of states of the
band fermions, Fig.~\ref{fig:dens}, to numerically evaluate the
energy and the variational coefficients $a$ and $b$ of the
eigenstate with the lowest energy. In Fig.~\ref{fig:bin}, we plot
this energy for different values of the parameter $\gamma m$ and
bare isotropic exchange coupling, $J$.

As expected, the binding energy decreases with increasing magnetic
order $m$. We also estimate the total moment of the collective
state by using the fermion  spin operator, $\textbf{s} =(1/2)
\sum_{{\bf k},n}a_{n{\bf k}\alpha}^{\dagger}{\bm
\sigma}_{\alpha\beta} a_{n{\bf k}\beta}$, and plotting the net
moment $(\textbf{S}+\textbf{s})^{2}$ as a function of $m$ in
Fig.~\ref{fig:bin}. Here $\textbf{S}$ is the impurity spin
operator, and the unscreened case corresponds to
$\textbf{S}^2=3/4$. As the antiferromagnetic order increases, the
screening is less effective and the moment at the impurity site
increases. Note significant change from the full screening in the
absence of AFM order, $(\bm S +\bm\sigma)^2=0$ when $m=0$, to a
substantial unscreened moment at $\gamma m=0.01$. This results
from the extreme sensitivity of Kondo screening to the low energy
features of the conduction band.

\section{Density of states.} 
\begin{figure}[t]
  \includegraphics[height=0.6\columnwidth,angle=-90]{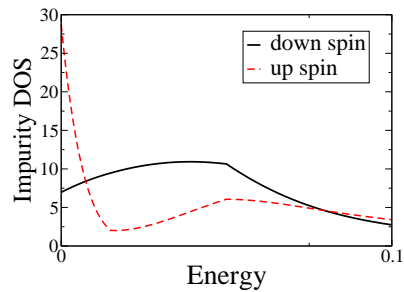}
  \caption{ Local spin-resolved DOS for the impurity located at the ``up-spin'' site
  for spins parallel(up) and antiparallel(down) to $m$ where  $\gamma m=0.05$ and $J=-0.04$.}
  \label{fig:prob}
\end{figure}

To estimate the single particle
density of states we compute $n(\epsilon,\sigma) = \left<\sum_{n} a_{\epsilon n\sigma}^{\dagger}a_{\epsilon n\sigma}\right>_{gs}$. The plot of the occupation for the two spins is shown in
Fig.~\ref{fig:prob}. Since the impurity sits at a site with the full symmetry of the square lattice, $m$ defines the magnitude and direction of the electronic moment at the site in the absence
of the impurity. Fig.~\ref{fig:prob} shows that the collective state formed by the impurity and the fermions leads to a splitting of the impurity density of states. The spin state antiparallel
to $m$ hybridizes with the fermions to give a rather broad peak shifted away from the chemical potential, while for the spins parallel to the sublattice magnetization, the density of states
shows two peaks, one at the chemical potential while the other at the antiferromagnetic energy scale $m$. We expect that the spin-resolved local density of states in the antiferromagnetic state
shows the same salient features even if computed fully self-consistently.

Our calculations have been performed in the magnetically ordered phase. We ignored the spin wave contribution, which becomes important near the magnetic phase transition; the effect of critical
fluctuations leads to a multichannel Kondo problem near the critical point \cite{MMV}. Coupling to the spin waves introduces a term of order $J^{2}g\chi$ to one loop where $g$ is the coupling of
order $\gamma m$ and $\chi$ the spin susceptibility  of the form $1/(\omega^{2}+\left|\textbf{q}-\textbf{Q}\right|^{2}+\xi^{-2})$. Since in the ordered phase $\xi$ is finite  for the modes that
couple to the fermions in the long-wavelength limit, the contributions due to spin waves are higher order in perturbation theory ($O( \gamma m J^{2} \xi^{2})$), and can be safely ignored here.

\section{Summary.}
To summarize we find that in an itinerant antiferromagnet, the
Kondo screening of the impurity moment competes with a spin
non-conserving coupling, which originates from the spin structure
of the quasiparticles eigenstates in the AFM phase, and the nature
of which depends on the local  symmetry of the impurity site. Our
result implies that in the heavy fermion AFM state the Kondo
screening is incomplete; this is in contrast to a very recent
variational Monte Carlo calculation for the Kondo lattice that
found screening of the moment across the transition from
paramegnet to antiferromagnet \cite{HWatanabe:2007}. We have
calculated by a simple variational method the residual ground
state moment and its distribution as a function of energy which
mimics what occurs in the antiferromagnetic heavy fermion state.
This is the first step in understanding the ordered phase of
Lattice Kondo model and its approach to criticality.

\section{Acknowledgements.} This research was supported in part by
Louisiana Board of Regents (I. V.) and by funds from the Los
Alamos-University of California, Riverside Joint Research Program
(V.A. and C. V.). We are grateful for the hospitality of Aspen
Center for Physics, where part of this work was done


\end{document}